\documentclass[12pt,onecolumn,oneside,letterpaper,peerreview]{IEEEtran}
\usepackage{cite}
\usepackage{graphicx}
\usepackage{amsmath}
\usepackage{times}
\usepackage{latexsym}
\usepackage{bm}
\usepackage{amssymb}
\usepackage[center]{caption2}
\usepackage{array}
\usepackage{setspace}
\usepackage{fancyhdr}
\usepackage{citesort}

\newtheorem{theorem}{Theorem}
\newtheorem{lemma}{Lemma}

\newcommand{\lj}{{\lambda_j}}

\newcommand{\defeq}{\stackrel{\Delta}{=}}

\newcaptionstyle{mystyle2}{%
\captionlabel $.$ \, \doublespacing \captiontext \par}
\captionstyle{mystyle2}

\begin{document}
\title{MIMO Multichannel Beamforming: SER and Outage Using New Eigenvalue Distributions of Complex Noncentral Wishart Matrices}
\author{
\begin{minipage}{0.98\columnwidth}
\vspace*{1cm}
\begin{center}
\authorblockN{Shi Jin\authorrefmark{2},
              Matthew R. McKay\authorrefmark{1}\authorrefmark{3},
              Xiqi Gao\authorrefmark{2},
              and Iain B. Collings\authorrefmark{3}\\
\vspace*{0.5cm}
\small{
\authorblockA{\authorrefmark{2}National Mobile Communications Research Laboratory,
Southeast University, Nanjing, China}\\
\authorblockA{\authorrefmark{1}Telecommunications Laboratory, Sch. of Elec. and Info.
Eng., University of Sydney, Australia} \\
\authorblockA{\authorrefmark{3}Wireless Technologies Laboratory, ICT Centre, CSIRO, Australia} } }
\end{center}
\end{minipage}
} \IEEEaftertitletext{\vspace{-0.75\baselineskip}}
\maketitle \vspace*{1cm}
\begin{abstract}
This paper analyzes MIMO systems with multichannel beamforming in
Ricean fading. Our results apply to a wide class of multichannel
systems which transmit on the eigenmodes of the MIMO channel. We
first present new closed-form expressions for the marginal ordered
eigenvalue distributions of complex noncentral Wishart matrices.
These are used to characterize the statistics of the signal to
noise ratio (SNR) on each eigenmode. Based on this, we present
exact symbol error rate (SER) expressions. We also derive
closed-form expressions for the diversity order, array gain, and
outage probability.  We show that the global SER performance is
dominated by the subchannel corresponding to the minimum channel
singular value. We also show that, at low outage levels, the
outage probability varies inversely with the Ricean $K$-factor for
cases where transmission is only on the most dominant subchannel
(i.e.\ a singlechannel beamforming system). Numerical results are
presented to validate the theoretical analysis.
\end{abstract}


\vskip 2ex \vspace*{0.5cm}
\begin{tabular}{ll}
{Corresponding Author\;:} {Shi Jin}\\
{National Mobile Communications Research Laboratory,}\\
{Southeast University, Nanjing 210096, China}  \\
{E-mail: jinshi@seu.edu.cn} \\
\end{tabular}

\newpage
\section{Introduction}\label{sec:introduction}

Multiple-input multiple-output (MIMO) communication systems have
received considerable attention in recent years as they offer
substantial capacity improvements over conventional single-input
single-output (SISO) systems \cite{Telatar,Foschini_limit}, with
no penalty in either power or bandwidth.  When perfect channel
state information (CSI) is available at both the transmitter and
receiver, it is well known \cite{Telatar} that the
capacity-achieving strategy is to transmit on the eigenmodes of
the MIMO channel using linear transmit-receive processing
(hereafter referred to as multichannel beamforming (MB)), upon
which independent Gaussian codes with water-filling power
allocation are employed.  In practice, the high complexity
requirements of Gaussian-like codes are often prohibitive,
%
%
and either suboptimally-coded or uncoded transmission is used.
Interestingly, for these suboptimal systems, it has been shown that
under most performance criteria of practical interest (e.g.\ symbol
error rate (SER), mean-square error (MSE), among others), MB still
corresponds to the optimal choice of linear transmit-receive
processing (in some cases up to a rotation matrix) \cite{Palomar03}.

In this paper we provide an analytical investigation of the
performance of uncoded MB MIMO systems in Ricean fading channels.
Although these systems are particularly appealing from a practical
point of view, there are currently very few related analytical
performance results available in the literature. In \cite{Garth},
the global SER (i.e.\ SER averaged over all subchannels) was derived
for Rayleigh and Ricean fading channels under the assumption that
all available subchannels were used for transmission, with equal
power, and with the same (BPSK) modulation. It was shown in
\cite{Palomar05} however (via Monte Carlo simulations), that the
performance is significantly improved by transmitting on only a
\emph{subset} of the available subchannels, and also by using
different powers and constellations on each of these. The global SER
results of\cite{Garth} cannot be easily generalized to these
important scenarios. In \cite{Luis}, global and per-subchannel SER
expressions were presented for MB MIMO systems in Rayleigh fading in
the high signal to noise ratio (SNR) regime, allowing for
transmission on a set of arbitrarily selected subchannels.  In this
paper we consider the more general class of Ricean fading channels,
and seek performance measures for all SNRs.

The main difficulty in obtaining analytical performance results for
MB MIMO systems in Ricean fading is that it requires the marginal
statistical distributions of the ordered eigenvalues of complex
noncentral Wishart matrices.  Although many results are available
for the eigenvalue statistics of complex \emph{central} Wishart
matrices (see
\cite{Burel,Edelman,Ratnarajah1,Zanella,McKay,Khatri1,Tulino04}, and
references therein), there are very few results for the noncentral
case.
In \cite{James}, the joint probability density function (p.d.f.) of
these ordered eigenvalues was obtained for the special case of
full-rank non-centrality matrices. This joint p.d.f.\ was used in
\cite{Smith} to derive the exact p.d.f.\ and cumulative distribution
function (c.d.f.) of the dual-antenna MIMO capacity in Ricean
channels with rank-$1$ mean matrices.  Unfortunately the joint
p.d.f.\ involves a hypergeometric function of matrix arguments and
Vandermonde determinants, and is not easily marginalized. In
\cite{Kang}, the p.d.f.\ and c.d.f.\ for the \emph{maximum}
eigenvalue was obtained for the special cases of rank-$1$ and
full-rank non-centrality matrices. These results were used to
analyze the statistics of the output SNR of MIMO maximum-ratio
combining (MIMO-MRC) systems in Ricean channels.

In this paper we derive new exact closed-form expressions for the
marginal distributions of \emph{all} of the ordered eigenvalues of
complex noncentral Wishart matrices.  The results apply for
non-centrality matrices of \emph{arbitrary} rank. Explicit
expressions are given for the marginal c.d.f.s, from which the
closed-form marginal p.d.f.s can be obtained trivially via
differentiation. These marginal c.d.f.s are used to analyze the
performance of MB MIMO systems in Ricean channels with mean matrices
of arbitrary rank. New exact expressions are presented for both the
subchannel SERs and the global SER. These expressions are general in
the sense that they allow for transmission on any arbitrary number
of subchannels, with possibly unequal signal constellations, and
with possibly unequal powers\footnote{Note that while we allow for
unequal powers, we assume that they remain fixed; and we calculate
the average SER.  In other words, we are not solving the
waterfilling problem, for which the power levels become functions of
the eigenvalues and change at the fading rate of the channel.}. The
exact c.d.f.\ expressions are also used to obtain new exact
closed-form expressions for the outage probability of MB MIMO
systems.

In addition to the exact marginal distributions, we also derive new
first-order asymptotic expansions for the marginal c.d.f.s and
p.d.f.s of the ordered eigenvalues of complex noncentral Wishart
matrices. These expansions are particularly useful for gaining
further insights into the effect of various system parameters on the
performance of MB MIMO systems. 
In particular, the asymptotic p.d.f.\ expansions are used to derive
explicit expressions for the diversity order and array gain,
which are the factors governing the SER performance at high SNR.
These expressions reveal that the global SER is dominated by the
subchannel corresponding to the minimum channel singular value. The
asymptotic c.d.f.\ expansions are used to examine the outage
probability of MB MIMO at low (practical) outage levels, when only a
single subchannel is selected for transmission (corresponding to
MIMO-MRC transmission). We find that in this outage regime, the
outage probability becomes independent of the rank of the channel
mean, and varies inversely with the Ricean $K$-factor.



\section{System Model}\label{sec:model}

Consider a MIMO system with $m$ transmit and $n$ receive antennas,
modeled as
\begin{equation}\label{CHANNEL_MODEL}
{\mathbf{y}} = {\mathbf{Hs}} + {\mathbf{n}},
\end{equation}
where ${\mathbf{y}} \in \mathbb{C}^{n \times 1}$ is the
discrete-time received vector, ${\mathbf{s}} \in \mathbb{C}^{m
\times 1}$ is the transmitted vector,
and ${\mathbf{n}} \in \mathbb{C}^{n \times 1}$ is the noise vector
with i.i.d.\ entries $\sim \mathcal{CN}(0,1)$.
Also, ${\mathbf{H}} \in \mathbb{C}^{n \times m}$ is the Ricean
fading channel matrix, which is decomposed as follows 
\cite{Farrokhi}
\begin{equation}\label{RICEAN_MODEL}
{\mathbf{H}} = \varepsilon \sqrt K {\mathbf{\bar H}} + \varepsilon
{\mathbf{\tilde H}},
\end{equation}
where ${\bf{\bar H}}$ is the (arbitrary) rank-$L$ deterministic
channel component satisfying ${\rm tr} \left( \mathbf{\bar{H}}
\mathbf{\bar{H}}^\dagger \right) = m n$, 
and ${\bf{\tilde H}}$ is the random (scattered) channel component
containing i.i.d.\ $\mathcal{CN}(0,1)$ entries. The parameter $K$
is the Ricean $K$-factor, which is the ratio between the energy in
${\bf{\bar H}}$ and the average energy in ${\bf{\tilde H}}$, and
$\varepsilon = 1/\sqrt {K + 1}$ is a power normalization constant.
Note that $\mathbf{H}$ in (\ref{RICEAN_MODEL}) follows a complex
matrix-variate Gaussian distribution with mean matrix ${\bf M} =
\sqrt K \varepsilon {\mathbf{\bar H}}$ and (column) correlation
matrix $\varepsilon^2 \mathbf{I}_n$. Adopting standard notation
from multivariate statistical theory (e.g.\ see
\cite{Muirhead,Gupta}), this distribution is denoted
\begin{align}
{\bf{H}} \sim \mathcal{CN}_{n,m} \left( {\bf M}, \, {\varepsilon^2
\mathbf{I}_n} \otimes {\bf{I}}_m \right).
\end{align}
Let us now define
\begin{equation}\label{WISHART}
{\mathbf{W}} = \left\{ \begin{array}{ll}
{\mathbf{H}} {\mathbf{H}}^\dag & \textrm{$n \leqslant m$}\\
{\mathbf{H}}^\dag  {\mathbf{H}}  & \textrm{$n > m$}\\
\end{array} \right.
\end{equation}
$s = \min(n,m)$, and $t = \max(n,m)$.  With these definitions,
${\mathbf{W}} \in \mathbb{C}^{s \times s}$ follows a complex
noncentral Wishart distribution, denoted
\begin{align} \label{eq:Wdefn}
{\mathbf{W}} \sim {\mathbf{W}}_s \left( {t,{\mathbf{\Sigma
}},{\mathbf{\Omega }}} \right),
\end{align}
where ${\bf \Sigma} = \varepsilon^2 \mathbf{I}_s$ and
\begin{align}\label{eq:omega}
{\bf{\Omega }} = \left\{ {\begin{array}{*{20}c}
   {{\bf{\Sigma }}^{ - 1} {\bf{MM}}^\dag  } & {n \leqslant m}  \\
   {{\bf{\Sigma }}^{ - 1} {\bf{M}}^\dag  {\bf{M}}} & {n > m}  \\
\end{array}} \right.
\end{align}
is the arbitrary-rank non-centrality matrix.

We consider the class of MB MIMO spatial multiplexing systems
considered in \cite{Palomar03,Luis,Palomar05}.  As in
\cite{Palomar03,Luis,Palomar05}, we assume that perfect CSI is known
at both the transmitter and receiver. 
The transmit vector can be written as
\begin{equation}\label{eq:transvec}
{\mathbf{s}} = {\mathbf{Bx}},
\end{equation}
where $\mathbf{B} \in \mathbb{C}^{m \times r}$ is the transmit
precoder matrix which maps the $r \leqslant \min \left( {m,n}
\right)$ modulated data symbols $x_i$ (elements of $\mathbf{x}$,
with $E\left\{\mathbf{x} \mathbf{x}^\dagger \right\} =
\mathbf{I}_r$, and chosen from possibly different signal
constellations), onto the $m$ transmit antennas, and is normalized
according to
\begin{align}
E \left\{ \| \mathbf{s} \|^2 \right\} = {\rm tr} \left( \mathbf{B}
\mathbf{B}^\dagger \right) \leqslant P
\end{align}
where $P$ is the average signal to noise ratio (SNR) per receive
antenna.  The estimated vector at the receiver is given by
\begin{equation}\label{eq:estimatvec}
{\mathbf{\hat x}} = {\mathbf{A}}^\dag  {\mathbf{y}},
\end{equation}
where ${\mathbf{A}}^\dag   \in \mathbb{C}^{r \times n}$ is the
receive (spatial) equalizer matrix.

It was shown in \cite{Palomar03}, that under many practical design
criteria (such as maximizing the mutual information, minimizing the
arithmetic or geometric mean-square error, among others) the optimal
transmit and receive filters result in a MB system, and are given by
\begin{align}
\mathbf{B} = \mathbf{U}_H \mathbf{P}
\end{align}
and
\begin{align}
\mathbf{A} = \left( \mathbf{H} \mathbf{B} \mathbf{B}^\dagger
\mathbf{H}^\dagger + \mathbf{I}_n \right)^{-1} \mathbf{H}\mathbf{B}
\end{align}
respectively, where $\mathbf{U}_H \in \mathbb{C}^{n \times r}$ has
as columns the eigenvectors corresponding to the $r$ largest
eigenvalues of $\mathbf{H}^\dagger \mathbf{H}$, and $\mathbf{P} =
{\rm diag} \left( \left\{ \sqrt{p_i} \right\}_{i=1, \ldots, r}
\right)$ is a power allocation matrix, with $p_i > 0$ and $\sum_i
p_i = P$.

With this choice of linear transmit and receive filtering, the MIMO
channel is decomposed into $r$ parallel scalar (eigenmode)
subchannels, which are described as follows
\begin{align}
\hat{x}_k = \kappa_k \left( \varepsilon \sqrt{ \phi_k p_k } x_k +
n_k\right)\, , \hspace*{1cm} k = 1, \ldots r
\end{align}
where $\kappa_k$ is a constant (which does not affect the received
subchannel SNR), $\hat{x}_k$ and $n_k$ are the $k$th elements of
$\mathbf{\hat x}$ and $\mathbf{n}$ respectively, and $\phi_k$ is the
$k$th largest eigenvalue of
\begin{align} \label{eq:sDefn}
\mathbf{S} = \mathbf{\Sigma}^{-1} \mathbf{W} \;  \sim \mathbf{W}_s
\left(t, \mathbf{I}_s, \mathbf{\Omega} \right) .
\end{align}
The instantaneous received subchannel SNR is given by
\begin{align} \label{eq:SNR_subchannel_inst}
\gamma_k = \varepsilon^2 \, \phi_k \, p_k \;  \hspace*{1cm} k = 1,
\ldots, r
\end{align}
Clearly the SNR (and therefore the performance) for each subchannel,
as well as the overall received SNR (and global performance), depend
explicitly on the marginal statistical distributions of the
eigenvalues $\phi_1 > \ldots > \phi_r$ of the complex noncentral
Wishart matrix in (\ref{eq:sDefn}).  In the following section we
will present new closed-form exact and asymptotic expressions for
the marginal distributions of these eigenvalues. These results will
then be used in Section \ref{sec:application} to analyze the
performance of MB MIMO systems in Ricean fading channels.

\section{New Statistical Properties of the Ordered Eigenvalues of Complex Noncentral Wishart Matrices}\label{sec:distributions}
\subsection{New Exact Ordered Eigenvalue Distribution Results}

In this subsection we derive new \emph{exact} closed-form marginal
eigenvalue c.d.f.\ expressions.  Note that exact marginal p.d.f.\
expressions can also be easily obtained from these c.d.f.\ results
via differentiation. These results, however, are omitted due to
space constraints. For convenience, we consider the smallest,
largest, and $k$th largest eigenvalues separately. The proofs of all
results in this section are given in the appendices.

First consider the smallest eigenvalue ${\phi _s }$.
It should be noted that, in additional to the performance analysis
of MB MIMO systems considered in this paper, the statistical
properties of the smallest eigenvalue of Wishart matrices are
important in the analysis of various other MIMO systems and
applications (see e.g.\ \cite{Narasimhan,Peel,Heath,Heath1}).
\begin{theorem}\label{smallest_cdf}
The c.d.f.\ of the smallest eigenvalue ${\phi _s }$ of the complex
noncentral Wishart matrix ${\mathbf{S}}$ in (\ref{eq:sDefn})
is given by
\begin{equation}\label{cdf_express}
F_{\phi _s } \left( x \right) = 1 - \left| {{\mathbf{\Psi }}\left(
x \right)} \right|/\left| {{\mathbf{\Psi }}\left( 0 \right)}
\right|,
\end{equation}
where ${{\mathbf{\Psi }} \left( x \right)}$ is an $s \times s$
matrix function of $x \in \left( {0, \infty } \right)$ whose entries
are given by
\begin{equation}\label{phiij_rc}
\left\{ {{\mathbf{\Psi }}\left( x \right)} \right\}_{i,j}  =
\left\{ {\begin{array}{*{20}c}
   {2^{\left( {2i - s - t} \right)/2} Q_{s + t - 2i + 1,t - s}
   \left( {\sqrt {2\lambda _j } ,\sqrt {2x} } \right)} & {j = 1, \ldots ,L}  \\
   {\Gamma \left( {t + s - i - j + 1,x} \right)} & {j = L + 1, \ldots ,s}  \\
 \end{array} } \right.
\end{equation}
and where $\lambda_1 > \ldots > \lambda_L$ are the $L \;
(\leqslant s)$ non-zero eigenvalues of $\mathbf{\Omega}$. Also,
$Q_{p,q} \left( {a,b} \right)$ is the Nuttall \textit{Q}-function,
defined in \cite{Nuttall} by
\begin{equation}\label{nuttall_q}
Q_{p,q} \left( {a,b} \right) = \int_b^{  \infty } {x^p \exp \left( {
- \frac{{x^2  + a^2 }} {2}} \right)I_q \left( {ax} \right) {\rm d}
x},
\end{equation}
$I_q \left(  \cdot  \right)$ is the $q$th order modified Bessel
function of the first kind, and $\Gamma \left( { \cdot , \cdot }
\right)$ is the upper incomplete gamma function, defined as
\cite{Gradshteyn}
\begin{equation}\label{ingamma_function_Up}
\Gamma \left( {p,x} \right)= \int_x^\infty  {t^{p - 1} e^{ - t} {\rm
d} t} = (p-1)! \, e^{-x} \sum_{k=0}^{p-1} \frac{ x^k}{k!},
\hspace*{1cm} p = 1, 2, \ldots.
\end{equation}
\end{theorem}
\begin{proof}
See Appendix \ref{sec:Proof_cdfRc}.
\end{proof}

Note that the result in (\ref{cdf_express}) can be easily programmed
and efficiently evaluated.  Moreover, the sum of the Nuttall
\textit{Q}-function indices in (\ref{phiij_rc}) is \emph{odd}.  As
such, this function has a closed-form representation given in
\cite[Eq.\ 8]{Simon}.

Now consider the largest eigenvalue $\phi_1$ of $\mathbf{S}$. In
\cite{Kang}, the c.d.f.\ of $\phi_1$ was derived for the particular
cases where the mean matrix was either rank-1 or full-rank, and used
to analyze the performance of MIMO-MRC systems in Ricean channels.
The following theorem presents a new, simpler expression for this
c.d.f., and also generalizes the results of \cite{Kang} as it
applies for mean matrices with \emph{arbitrary} rank.
\begin{theorem}\label{largest_cdf}
The c.d.f.\ of the largest eigenvalue ${\phi _1 }$ of the complex
noncentral Wishart matrix ${\mathbf{S}}$ in (\ref{eq:sDefn}) is
given by
\begin{equation}\label{largest_cdfexp}
F_{\phi _1 } \left( x \right) = \left| {{\mathbf{\Xi }}\left( x
\right)} \right|/\left| {{\mathbf{\Psi }}\left( 0 \right)}
\right|,
\end{equation}
where ${{\mathbf{\Xi }}\left( x \right)}$ is an $s \times s$ matrix
function of $x \in \left( {0, \infty } \right)$ whose entries are
given by
\begin{equation}\label{Xi_exp}
\left\{ {{\mathbf{\Xi }}\left( x \right)} \right\}_{i,j}  =
\left\{ {\begin{array}{*{20}c}
   {2^{\left( {2i - s - t} \right)/2} \left[ {Q_{s + t - 2i + 1,t - s}
   \left( {\sqrt {2\lambda _j } ,0} \right) - Q_{s + t - 2i + 1,t - s}
   \left( {\sqrt {2\lambda _j } ,\sqrt {2x}} \right)} \right]}
   & {j = 1, \ldots ,L}  \\
   {\gamma \left( {t + s - i - j + 1,x} \right)}
   & {j = L + 1, \ldots ,s}  \\
 \end{array} } \right.
\end{equation}
where $\gamma(\cdot, \cdot)$ is the lower incomplete gamma
function, given by
\begin{equation}\label{ingamma_function_Low}
\gamma \left( {p,x} \right)  = \int_0^x {t^{p - 1} e^{ - t} {\rm d}
t} = (p-1)! \, \left(1 - e^{-x} \sum_{k=0}^{p-1} \frac{ x^k}{k!}
\right), \hspace*{1cm} p = 1, 2, \ldots.
\end{equation}
\end{theorem}
\begin{proof}
See Appendix \ref{sec:Proof_largest}.
\end{proof}

Finally we consider the $k$th largest eigenvalue $\phi_k$ of
$\mathbf{S}$.
\begin{theorem}\label{thekth_cdf}
The c.d.f.\ of the $k$th largest eigenvalue ${\phi _k }$ of the
complex noncentral Wishart matrix ${\mathbf{S}}$ in
(\ref{eq:sDefn}), where $k>1$, is given by
%
\begin{align}\label{thekthcdf_exp}
F_{\phi _k } \left( x \right) &= F_{\phi _{k-1} } \left( x \right)
+ \Pr \left( {\phi _s  <  \ldots  < x < \phi _{k - 1}  <  \ldots
< \phi _1 } \right) = F_{\phi _{k-1} } \left( x \right) + p,
\end{align}
where
\begin{equation}\label{thekthcdf_pexp}
p = c_3 \sum\limits_1 {\left| {\mathbf{\Theta }}\left( x \right)
\right|},
\end{equation}
\begin{equation}\label{c3_expression}
c_3  = \frac{{\prod\nolimits_{i = 1}^L {\left( {\lambda _i^{\left(
{2L - s - t} \right)/2} } \right)} }} {{\Gamma _{s - L} \left( {s
- L} \right)\prod\nolimits_{i < j}^L {\left( {\lambda _i  -
\lambda _j } \right)} }},
\end{equation}
\begin{equation}\label{Zk_exp}
\left\{ {{\mathbf{\Theta }}\left( x \right)} \right\}_{\alpha _i
,j}  = \left\{ {\begin{array}{*{20}c}
   {\left\{ {{\mathbf{\Psi }}\left( x \right)} \right\}_{\alpha _i ,j} } & {i = 1, \ldots ,k - 1}  \\
   {\left\{ {{\mathbf{\Xi }}\left( x \right)} \right\}_{\alpha _i ,j} } & {i = k, \ldots ,s}  \\
 \end{array} } \right.
\end{equation}
$\sum\limits_1$ indicates the summation over the combination
$\left( {\alpha _1  < \alpha _2  <  \ldots  < \alpha _{k - 1} }
\right)$ and $\left( {\alpha _k  < \alpha _{k + 1}  <  \ldots  <
\alpha _s } \right)$, $\left( {\alpha _1 , \ldots ,\alpha _s }
\right)$ being a permutation of $\left( {1, \ldots ,s} \right)$.
\end{theorem}
\begin{proof}
See Appendix \ref{sec:Proof_thekthRc}.
\end{proof}

\textit{Remark:} Let $\omega _1  \geqslant \omega _2  \geqslant
\cdots \geqslant \omega _s \geqslant 0$ be the ordered singular
values of ${\mathbf{H}}$. Recalling that ${\bf{S}} = {\bf{\Sigma
}}^{ - 1} {\bf{W}}$, we have the following relationship
\begin{equation}\label{relation_HS}
F_{\omega _s } \left( x \right) = F_{\phi _s } \left( {\varepsilon
^{ - 2} x^2 } \right).
\end{equation}

\subsection{New Asymptotic Ordered Eigenvalue Distribution Results}

In this subsection we present new asymptotic \emph{first-order
expansions} of the marginal eigenvalue p.d.f.s and c.d.f.s.\ of
complex noncentral Wishart matrices. These will be particularly
useful for deriving the diversity order and array gain of MB MIMO
systems in the following section, as well as for analyzing the
asymptotic outage probability. Note that a first order expansion of
the p.d.f.\ of the $k$th largest eigenvalue of complex
\emph{central} Wishart matrices was presented in \cite{Luis}.
\begin{theorem}\label{pdf_approximation}
The first order expansions of the marginal p.d.f.\ and c.d.f.\ of
the $k$th largest eigenvalue $\phi_k$ of the complex noncentral
Wishart matrix ${\mathbf{S}}$ in (\ref{eq:sDefn}), where $1
\leqslant k \leqslant s$, are given respectively by
%
\begin{equation}\label{first_order1}
f_{\phi _k } \left( {\phi _k } \right) = a_k \phi _k^{d_k }  +
o\left( {\phi _k^{d_k } } \right)
\end{equation}
and
\begin{equation}\label{first_order1_cdf}
F_{\phi _k } \left( {\phi _k } \right) = \frac{a_k}{d_k + 1} \phi
_k^{d_k + 1 } + o\left( {\phi _k^{d_k + 1 } } \right),
\end{equation}
where
\begin{equation}\label{dk_totalexp}
d_k  = s_k t_k - 1
\end{equation}
and
\begin{align}
s_k = s - k + 1, \hspace*{1cm} t_k = t-k+1 .
\end{align}
Also, $a_k$ is given for $k=1$ by
\begin{align}
a_1 = \frac{ s t \, \Gamma_s (s)}{ \Gamma_s (t+s)} \, e^{-{\rm tr}(
\mathbf{\Omega} )}
\end{align}
and for $k > 1$ by
\begin{align}
a_k = \frac{ s_k t_k \Gamma_{k-1}(s) \Gamma_{s_k}(s_k)
}{\Gamma_{s_k}(t_k+s_k)} c_3 \left(\prod\nolimits_{i = 1}^L {\lambda
_i^{\left( {s - t} \right)/2} } \right) \left| {\bf{X}} \right|
\end{align}
where $\Gamma_\cdot (\cdot)$ is the normalized complex multivariate
gamma function, defined in (\ref{eq:mvgam}), and $\mathbf{X}$ is an
$s \times s$ matrix with $(i,j)$th entry
\begin{align}
\left\{ {\mathbf{X}} \right\}_{i,j} = \left\{
\begin{array}{ll}
L_{s-i}^{(t-s)} (-\lambda_j)  \hspace*{1cm} & \textrm{$i = 1, \ldots, k-1, \; \; \; \; j = 1, \ldots, L$}\\
\lambda_j^{s-i} e^{-\lambda_j} \hspace*{1cm} & \textrm{$i = k, \ldots, s, \; \; \; \; j = 1, \ldots, L$}\\
\binom{t-i}{j-i}
\hspace*{1cm} & \textrm{$i = 1, \ldots, k-1, \; \; \; \; j = L+1, \ldots, s, \; \; \; j \geqslant i$}\\
0 \hspace*{1cm} & \textrm{$i = 1, \ldots, k-1, \; \; \; \; j = L+1, \ldots, s, \; \; \; j < i$}\\
\frac{(-1)^{i-j} (s-j)!}{(i-j)!}
\hspace*{1cm} & \textrm{$i = k, \ldots, s, \; \; \; \; j = L+1, \ldots, s, \; \; \; j \leqslant i$}\\
0
\hspace*{1cm} & \textrm{$i = k, \ldots, s, \; \; \; \; j = L+1, \ldots, s, \; \; \; j > i$}\\
%
\end{array} \right.
\end{align}
where
\begin{equation}\label{laguerre_k}
L_k^{\left( n \right)} \left( x \right) = \sum\limits_{i = 0}^k
\binom{k+n}{k-i} 
\frac{{\left( { - x} \right)^i }} {{i!}}
\end{equation}
is the generalized $k$th-order Laguerre polynomial.
\end{theorem}
\begin{proof}
See Appendix \ref{sec:Proof_pdfappro}.
\end{proof}

\section{Performance Analysis of MIMO Multichannel Beamforming in Ricean Fading Channels}\label{sec:application}
\subsection{Symbol Error Rate Analysis}\label{sec:sepsvdmimo}

We now analyze the SER performance of the MB MIMO systems introduced
in Section \ref{sec:model}. For many general modulation formats (see
below), the average SER of the $k$th subchannel can be expressed as
\cite{Proakis01}
\begin{equation}\label{ber_average1}
{\rm SER}_{k}  = E_{\gamma _k } \left\{ {\alpha_k Q\left( {\sqrt {2
\beta_k \gamma _k } } \right)} \right\}, \hspace*{1cm} k = 1,
\ldots, r
\end{equation}
where $Q(\cdot)$ is the Gaussian Q-function, and $\alpha_k$ and
$\beta_k$ are modulation-specific constants. Some example modulation
formats for which (\ref{ber_average1}) apply include BPSK $(\alpha_k
= 1, \, \beta_k = 1)$; BFSK with orthogonal signalling $(\alpha_k =
1, \, \beta_k = 0.5)$ or minimum correlation $(\alpha_k = 1,  \,
\beta_k = 0.715)$; and $M-$ary PAM $( \alpha_k = 2(M-1)/M,  \,
\beta_k = 3/(M^2-1))$. Our results also provide the approximate SER
for those other formats for which (\ref{ber_average1}) is an
approximation, e.g.\ $M$-ary PSK $(\alpha_k = 2, \beta_k = \sin^2 (
\pi/M ) )$ \cite[Eq. 5.2-61]{Proakis01}.
%
%
%
Using results from \cite{Zanella} and \cite{Chen04},
(\ref{ber_average1}) can be expressed in the following equivalent
form
\begin{equation}\label{ber_average2}
{\rm SER}_{k}  = \frac{{\alpha_k\sqrt \beta_k }} {{2\sqrt \pi
}}\int_0^\infty {\frac{{e^{ - \beta_k u} }} {{\sqrt u }}} F_{\phi _k
} \left( \frac{{u}} {{\varepsilon ^2 p_k }} \right){\rm d} u,
\hspace*{1cm} k = 1, \ldots, r
\end{equation}
where we have used the fact that $F_{\gamma _k } \left( u \right)
= F_{\phi _k } \left( \varepsilon ^{ - 2} u/p_k \right)$. Hence,
by applying \emph{Theorems \ref{smallest_cdf}-\ref{thekth_cdf}} in
(\ref{ber_average2}) we obtain an exact expression for the average
SER of each subchannel.  Although it does not appear that the
integrals in (\ref{ber_average2}) can be evaluated in closed form,
numerical integration can be performed to evaluate ${\rm SER}_{k}$
much more efficiently than is possible via Monte Carlo simulation.

The global SER can be derived from the subchannel SERs as follows;
since independent symbols are sent on each subchannel during each
channel use:
\begin{equation}\label{ser_svdMIMO}
{\rm SER}  = \frac{1} {r}\sum\limits_{k = 1}^r {{\rm SER}_{k} }.
\end{equation}

To gain further insights, we now consider the SER at high SNR. We
will restrict this asymptotic analysis to systems with uniform power
allocation (i.e.\ $p_k = P/r$), since (as mentioned in \cite{Luis})
most of the practical power allocation solutions in \cite{Palomar03}
tend to uniform as the power is increased.  In the high SNR regime,
the key factors governing system performance are the diversity order
and array gain. We will now present closed-form expressions for
these factors.

Using a general SISO SER result from \cite{Wang}, we find that in
our case ${\rm SER}_{k}$ can be approximated in the high SNR regime
by considering a first order expansion of the p.d.f.\ of $\phi_k$ as
$\phi_k \rightarrow 0^+$. Hence, using the result from \cite{Wang},
along with \emph{Theorem \ref{pdf_approximation}}, we obtain a high
SNR subchannel SER expression given by
\begin{equation}\label{SERsubk_approximation}
{\rm SER}_{k}^\infty = \left( {G_a \left( k \right) \cdot P }
\right)^{ - G_d \left( k \right)}  + o\left( {P ^{ - G_d \left( k
\right)} } \right), \hspace*{1cm} k = 1, \ldots, r
\end{equation}
where the diversity order is
\begin{equation}\label{gdk_exp}
G_d \left( k \right) = d_k  + 1
\end{equation}
and the array gain is
\begin{equation}\label{gak_exp}
G_a \left( k \right) = \frac{2 \, \beta_k \, \varepsilon}{r}
^2\left( {\frac{{\alpha_k 2^{d_k } a_k \Gamma \left( {d_k  + 3/2}
\right)}} {{\sqrt \pi  \left( {d_k + 1} \right)}}} \right)^{ -
1/\left( {d_k + 1} \right)}.
\end{equation}

Comparing (\ref{gdk_exp}) with the i.i.d.\ Rayleigh results
presented previously in \cite{Luis}, we see that the subchannel
diversity orders are the same in both Rayleigh and Ricean channels.
Moreover, since $G_d \left( 1 \right) \geqslant G_d \left( 2 \right)
\geqslant \cdots  \geqslant G_d \left( r\right)$, the $r$th
subchannel has the poorest performance in terms of average SER.
Using (\ref{ser_svdMIMO}), the global average SER of MB MIMO systems
at high SNR can be obtained as
\begin{equation}\label{global_SER1}
{\rm SER}^\infty = \frac{1} {r}\left( {G_a \left( r \right) \cdot P
} \right)^{ - G_d \left( r \right)}  + o\left( {P ^{ - G_d \left( r
\right)} } \right)
\end{equation}
which is clearly dominated by the $r$th subchannel SER (i.e.\ the
subchannel corresponding to the smallest singular value).

\subsection{Outage Probability Analysis}\label{sec:outsepmrcmimo}

We now consider the outage probability of MB MIMO systems in Ricean
fading channels. The outage probability is an important quality of
service measure, defined as the probability that the received SNR
drops below an acceptable SNR threshold $\gamma_{{\rm th}}$.  For
convenience, we assume an equal power allocation strategy is
employed.
In this case, the subchannel SNRs are ordered according to $\gamma_1
> \ldots > \gamma_r$ in (\ref{eq:SNR_subchannel_inst}), and the
outage probability of the overall MB MIMO system is dominated by the
subchannel corresponding to $k = r$.  As such, the outage
probability is obtained exactly from \emph{Theorems 1-3} as follows
\begin{equation}\label{MIMO_MRC_outage}
P_{{\rm out}}  = \Pr \left( {\gamma_r  \leqslant \gamma _{{\rm th}}
} \right) = F_{\phi _r } \left( {\frac{{\gamma _{{\rm th}} \left( {K
+ 1} \right) r}} {{P }}} \right) \; .
\end{equation}
We note that for the special case $r = 1$ (i.e.\ only the best
subchannel corresponding to $\phi_1$ is used), the MB MIMO system
we consider is equivalent to the MIMO-MRC systems considered in
\cite{Kang,McKay}. For these systems, outage probability
expressions were derived previously in \cite{Kang} for the special
case of channels with rank-1 and full-rank mean matrices.  Our
result (\ref{MIMO_MRC_outage}) is clearly more general as it
applies for all $r \geqslant 1$ and for mean matrices of arbitrary
rank.  Moreover, for the special case $r = 1$, our result is
simpler than the previous results given in \cite{Kang}.

In practice, we are usually interested in small outage probabilities
(i.e. 0.01, 0.001, ...), which correspond to small values of
$\gamma_{\rm th}$. To gain further intuition at these small outage
probabilities, let us consider the case $r = 1$ (i.e.\ the MIMO-MRC
case), and use (\ref{first_order1_cdf}) in \textit{Theorem
{\ref{pdf_approximation}}}
to write the outage probability in (\ref{MIMO_MRC_outage}) as
follows
\begin{equation}\label{MIMO_MRC_appoutage}
\tilde P_{{\rm out}}  = \frac{{\Gamma _s \left( s \right)e^{ - {\rm
tr}\left( {\mathbf{\Lambda }} \right)} }} {{\Gamma _s \left( {t + s}
\right)}}\left( {\frac{{\gamma _{{\rm th}} \left( {K + 1} \right)}}
{{P}}} \right)^{st}  + o\left( {\left( {\gamma _{\rm th} }
\right)^{st} } \right).
\end{equation}
Since ${\rm tr} \left( \mathbf{\bar{H}} \mathbf{\bar{H}}^\dagger
\right) = s t$ (see Section \ref{sec:model})
we can further simplify to obtain
\begin{equation}\label{MIMO_MRC_appoutage1}
\tilde P_{{\rm out}}  = \frac{{\Gamma _s \left( s \right)}} {{\Gamma
_s \left( {t + s} \right)}}\left( {\frac{{\gamma _{{\rm th}} }} {{P
}}} \right)^{st} \frac{{\left( {K + 1} \right)^{st} }} {{e^{{Kst}}
}} + o\left( {\left( {{\gamma _{\rm th} } } \right)^{st} } \right).
\end{equation}
This explicitly shows that for MIMO-MRC transmission the outage
probability (at low outage levels) does not depend on the rank of
the channel mean.  Moreover, since
\begin{align}
\frac{d} {{dK}}\left( {K + 1} \right)^{st} e^{ - Kst}  =  - Kste^{
- Kst} \left( {K + 1} \right)^{st - 1}  < 0,{\text{ }}K > 0
\end{align}
we see that the outage probability varies inversely with the
Ricean $K$-factor.


\section{Numerical Results}\label{sec:numerical}

For our numerical results we consider a $3 \times 5$ Ricean MIMO
channel and, unless otherwise specified, a rank-$3$ deterministic
component ${\mathbf{\bar H}}$ with singular values
$\left\{ 2.9751, \, 2.2840, \, 0.9657 \right\}$, which were randomly
generated to verify the analysis.

It is important to note, however, that all of the analytic results
in this paper apply for arbitrary antenna configurations, and for
arbitrary deterministic channel components.

\begin{figure}
\centering
\includegraphics[scale=0.7]{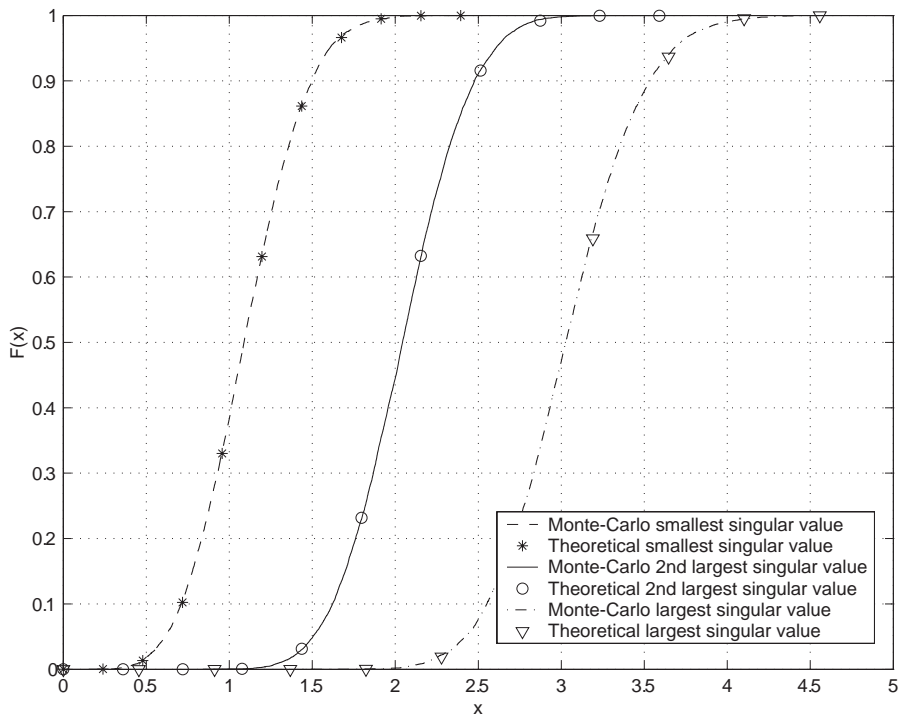}
\captionstyle{mystyle2}\caption{Ordered c.d.f.s of the singular
values of a $3 \times 5$ Ricean MIMO channel, with $K=10 {\rm
dB}$.} \label{fig:fig1}
\end{figure}

Fig.\ \ref{fig:fig1} gives the marginal c.d.f.s of the ordered
channel singular values. 
The analytical curves are
generated using \emph{Theorems 1-3} and the singular value
relationship (\ref{relation_HS}), and the simulated curves are
generated based on 100,000 channel realizations. The figure shows
perfect agreement between the analytical results and the
simulations.

\begin{figure}
\centering
\includegraphics[scale=0.7]{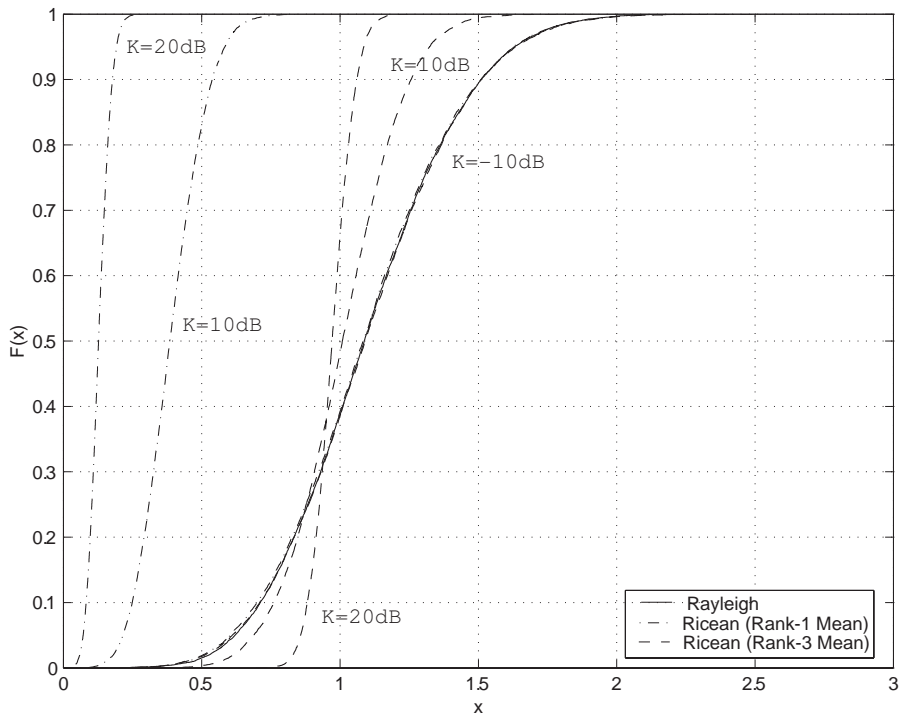}
\caption{Smallest singular value c.d.f.\ for a $3 \times 5$ Ricean
MIMO channel with rank-1 and rank-3 channel means, and for different
$K$-factors. Rayleigh c.d.f.\ presented for comparison.}
\label{fig:fig2}
\end{figure}

Fig.\ \ref{fig:fig2} shows the c.d.f.\ of the smallest singular
value for different Ricean $K$-factors, and for rank-$1$ and
rank-$3$ mean matrices. The deterministic component for the rank-$1$
case was generated based on the channel model in \cite{Smith}.
As expected, for both mean matrices, as $K$ becomes small (i.e.\ $K
= -10 {\rm dB}$), the c.d.f.s converge to that of the smallest
singular value of a Rayleigh channel.

\begin{figure}
\centering
\includegraphics[scale=0.7]{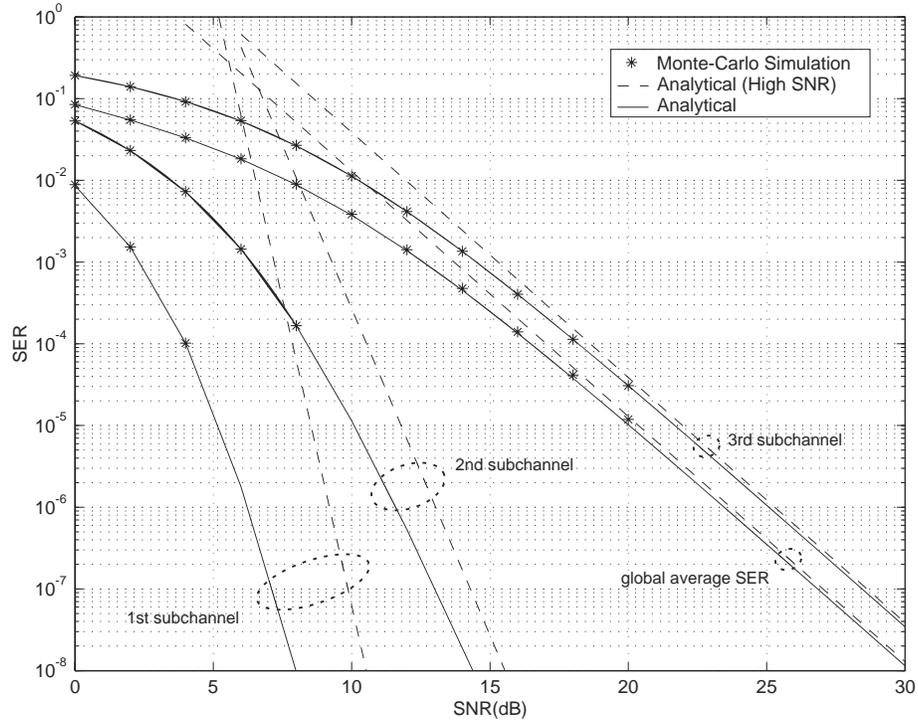}
\caption{Exact analytical SER, high SNR analytical SER, and
Monte-Carlo simulated SER for $3 \times 5$ MB MIMO in uncorrelated
Ricean fading, with rank-3 mean matrix and $K=0 {\rm dB}$.}
\label{fig:fig6}
\end{figure}

Fig.\ \ref{fig:fig6} shows the exact subchannel SER curves based on
(\ref{ber_average2}), exact global SER curve based on
(\ref{ser_svdMIMO}), and Monte-Carlo SER simulation results, for a
MB MIMO system with $K=0 {\rm dB}$. All subchannels are used with
BPSK modulation ($\alpha_k = 1$, $\beta_k = 1$) and uniform power
allocation. High SNR curves based on (\ref{SERsubk_approximation})
and (\ref{global_SER1}) are also presented.  In all cases, there is
exact agreement between the analytical SER results and the
Monte-Carlo simulations, and the diversity order and array gains
predicted by the high SNR analytical results are accurate. We also
see that the SERs of the $1$st and $2$nd subchannels are
\emph{significantly} better than the $3$rd subchannel SER. This
suggests that further performance improvements may be possible by
using only a subset of the subchannels for transmission (with higher
order constellations).  The following figure investigates this
further.

Fig.\ \ref{fig:fig5} shows the analytical and Monte-Carlo
simulated global SER curves for MB MIMO systems with different
numbers of $r$ active subchannels. The analytical curves are
generated based on (\ref{ser_svdMIMO}). Uniform power allocation
is assumed and, for a fair comparison, the overall rate is set to
$3$ bits/s/Hz in each case. For $r=1$, $8$PSK ($\alpha_1 = 2 $,
$\beta_1 = 0.146$) is employed; for $r = 2$ we use QPSK ($\alpha_1
= 2 $, $\beta_1 = 0.5$) for the first subchannel and BPSK for the
second subchannel; and for $r=3$ we use BPSK for all subchannels.
For the $r=3$ case see an exact agreement between the analytical
and Monte-Carlo simulated curves. As discussed in Section
\ref{sec:sepsvdmimo}, (\ref{ser_svdMIMO}) only provides an
approximation for QPSK and $8$PSK, however we see from the $r=2$
and $r=3$ curves that the approximation is accurate. In
particular, for low to moderate SERs (i.e.\ ${\rm SER} < 10^{-3}$)
the analytical curves match almost exactly with the simulated
curves.  We also see that the SER for the $r=1$ and $r=2$ cases is
significantly better than for the case where all subchannels are
used, which is in agreement with previous Rayleigh fading
observations in \cite{Palomar05} (via Monte-Carlo simulations).
Moreover, we see that the $r=1$ system has a higher diversity
order than the $r=2$ system (since we've shown the diversity order
to be dominated by lowest singular value subchannel), but is
shifted to the right due to the higher order constellations. This
motivates the design of practical adaptive subchannel selection
algorithms, which is an interesting topic for future work, but
beyond the scope of this paper.

Fig.\ \ref{fig:fig4} shows analytical and Monte-Carlo simulated
outage probability curves for a MB MIMO systems with $r=1$ (i.e.\
MIMO-MRC transmission), comparing different Ricean $K$-factors.
The analytical curves are generated based on
(\ref{MIMO_MRC_outage}). We see an exact agreement between the
analytical and simulated curves in all cases. We also see that
increasing the Ricean $K$-factor results in a reduction in outage
probability (and an improvement in system performance) at low
outage levels. This agrees with the analytic conclusions given in
Section \ref{sec:outsepmrcmimo}. It is also interesting to observe
that the opposite occurs in the high outage regime (i.e.\
increasing the $K$-factor increases the outage probability).

\section{Conclusion}\label{sec:conclusion}

We have examined the performance of MIMO systems employing
multichannel beamforming in arbitrary-rank Ricean channels. Our
results are based on new closed-form exact and asymptotic
expressions which we have derived for the marginal ordered
eigenvalue distributions of complex noncentral Wishart matrices.  We
have presented exact and high-SNR SER expressions, and derived the
diversity order and array gain. Our results have shown that the
global SER performance is dominated by the subchannel SER
corresponding to the minimum channel singular value. We have also
derived new closed-form exact expressions for the outage probability
and, for the case of MIMO-MRC transmission, have shown that for
outage levels of practical interest, the outage probability varies
inversely with the Ricean $K$-factor.

\appendices
\section{Proof of Theorem \ref{smallest_cdf}}\label{sec:Proof_cdfRc}

We require the following Lemma, which gives the joint eigenvalue
p.d.f.\ of $\mathbf{S}$ for the case where the non-centrality matrix
$\mathbf{\Omega}$ has \emph{arbitrary-rank}.

\begin{lemma}\label{lemma1}
The joint p.d.f.\ of the ordered eigenvalues $\phi _1
> \phi _2  > \ldots  > \phi _s > 0$ of the complex noncentral Wishart matrix ${\mathbf{S}}$ 
in (\ref{eq:sDefn}) is given by
\begin{equation}\label{order_pdf1}
f\left( {\phi _1 , \ldots ,\phi _s } \right) = c_1 {\left|
{\mathbf{\Upsilon }}  \right|}\prod\limits_{i < j}^s \left( {\phi _i
- \phi _j } \right) \prod\limits_{k = 1}^s {\phi _k^{t - s} } e^{ -
\phi _k },
\end{equation}
where
\begin{equation}\label{c1_expression}
c_1  = \frac{{e^{ - {\rm tr}\left( {\mathbf{\Omega }} \right)}
\left( {\left( {t - s} \right)!} \right)^{ - s} }} {{\Gamma _{s - L}
\left( {s - L} \right)\prod\nolimits_{i = 1}^L {\left( {\lambda
_i^{s - L} } \right)\prod\nolimits_{i < j}^L {\left( {\lambda _i -
\lambda _j } \right)} } }}
\end{equation}
and where
\begin{align} \label{eq:mvgam}
\Gamma_{s} (t) = \prod_{i=1}^{s} ( t - i)! \; .
\end{align}
Also
${\mathbf{\Upsilon }}$ is an $s \times s$ matrix with ($i$,$j$)th
entry
\begin{equation}\label{Upsilon_exp}
\left\{ \mathbf{\Upsilon }  \right\}_{i,j}  = \left\{
{\begin{array}{*{20}c}
   {{}_0F_1 \left( {t - s + 1;\lambda _j \phi _i } \right)} & {i = 1, \ldots ,s,} & {j = 1, \ldots ,L}  \\
   {\frac{{\phi _i^{s - j} \left( {t - s} \right)!}}
{{\left( {t - j} \right)!}}} & {i = 1, \ldots ,s,} & {j = L + 1, \ldots ,s}  \\
 \end{array} } \right.
\end{equation}
where ${}_0F_1(\cdot)$ is the scalar Bessel-type hypergeometric
function.
\end{lemma}
\begin{proof}
We start by combining a result from \cite{James}, which gave the
joint eigenvalue p.d.f.\ for the special case of \emph{full-rank}
$\mathbf{\Omega}$, with a hypergeometric function determinant result
from \cite{Gross}, to express the joint eigenvalue p.d.f.\ in the
full-rank $\mathbf{\Omega}$ case as follows
\begin{equation}\label{ordered_dis}
f_{\rm FR}\left( {\phi _1 , \ldots ,\phi _s } \right) = \frac{ e^{ -
{\rm tr}\left( {\mathbf{\Omega }} \right)}  \, e^{ - {\rm tr}\left(
{\mathbf{\Phi }} \right)}}{{\left( {\left( {t - s } \right)!}
\right)^s }} \prod\limits_{i < j}^s \left( {\phi _i - \phi _j }
\right) \prod\limits_{k = 1}^s {\phi _k^{t - s} } \frac{ \, {\left|
{{}_0F_1 \left( {t - s + 1;\lambda _j \phi _i } \right)} \right|}}
 { { \; \prod\nolimits_{i < j}^s {\left( {\lambda _i  -
\lambda _j } \right)} } }
\end{equation}
where ${\mathbf{\Lambda }} = {\rm diag}\left( {\lambda _1 , \ldots
,\lambda _s } \right)$, $\lambda _1  >  \ldots  > \lambda _s > 0$
are the eigenvalues of $\mathbf{\Omega }$, and ${\mathbf{\Phi }} =
{\rm diag}\left( {\phi _1 , \ldots ,\phi _s } \right)$. 
We generalize this result to
arbitrary-rank matrices $\mathbf{\Omega}$ by taking limits
\begin{align} \label{eq:jointpdfreducerank}
f\left( {\phi _1 , \ldots ,\phi _s } \right) &= \mathop {\lim
}\limits_{\lambda_{L+1} , \ldots ,\lambda _s  \to 0} f_{\rm
FR}\left( {\phi _1 , \ldots ,\phi _s } \right) = \frac{ e^{ - {\rm
tr}\left( {\mathbf{\Omega }} \right)}  \, e^{ - {\rm tr}\left(
{\mathbf{\Phi }} \right)}}{{\left( {\left( {t - s } \right)!}
\right)^s }} \prod\limits_{i < j}^s \left( {\phi _i - \phi _j }
\right) \prod\limits_{k = 1}^s {\phi _k^{t - s} }  \; \mathcal{L}
\end{align}
where
\begin{align} \label{eq:Ldefn}
\mathcal{L} = \mathop {\lim} \limits_{\lambda_{L+1} , \ldots
,\lambda _s  \to 0} \frac{\left| f_i \left( {\lambda_j } \right)
\right|}{{ \; \prod\nolimits_{i < j}^s {\left( {\lambda _i  -
\lambda _j } \right)} } } \,
\end{align}
with $f_i \left( {\lambda_j } \right) = {}_0F_1 \left( {t - s +
1;\lambda_j \phi _i } \right)$. To evaluate these limits we apply
\cite[Lemma 2]{Chiani06} to obtain
\begin{align}\label{df1_det_lim1}
\mathop {\lim }\limits_{\lambda _{L + 1} , \ldots ,\lambda _s  \to
0} \frac{{\left| {f_i \left( {\lambda _j } \right)} \right|}}
{\prod\nolimits_{i < j}^s {\left( {\lambda _i  - \lambda _j }
\right)}} &= \frac{\left|
\begin{matrix}
 f_1 \left( {\lambda _1 } \right) & \ldots & f_1 \left(
{\lambda _L } \right) & f_1^{\left( {s - L - 1} \right)} \left( 0
\right) & \ldots  & f_1^{(0)} \left( 0 \right) \\
\vdots & \vdots &  & & \vdots & \vdots \\
 f_s \left( {\lambda _1 } \right) & \ldots & f_s \left(
{\lambda _L } \right) & f_s^{\left( {s - L - 1} \right)} \left( 0
\right) & \ldots  & f_s^{(0)} \left( 0 \right)
\end{matrix} \right|}{{\prod\nolimits_{i < j}^L {\left( {\lambda _i
- \lambda _j } \right)}  {\left( {\prod\nolimits_{i = 1}^L {\lambda
_i^{s - L} } } \right)} \Gamma _{s - L} \left( {s - L} \right)}}
\end{align}
where the required derivatives are easily evaluated as
\begin{equation}\label{eta_lim}
f_i^{\left( k \right)} \left( 0 \right) = \frac{{\phi _i^k \left(
{t - s} \right)!}} {{\left( {t - s + k} \right)!}}.
\end{equation}
The result now follows by substituting
(\ref{eq:Ldefn})-(\ref{eta_lim}) into (\ref{eq:jointpdfreducerank})
and simplifying.
\end{proof}

We now proceed to evaluate the c.d.f.\ of the minimum eigenvalue
$\phi_s$ as follows
\begin{equation}\label{smallest_eig}
F_{\phi _s } \left( x \right) = 1 - \Pr \left( {\phi_1 > \ldots \phi
_s
> x} \right) = 1 - \int_{\mathcal{D}_1} f\left( {\phi _1 , \ldots
,\phi _s } \right) \, {\rm d}\phi_1 \cdots {\rm d}\phi_s \,
\end{equation}
where $D_1  = \left\{ {x < \phi_s  < \ldots  < \phi_1 } \right\}$.
To evaluate the integrals in (\ref{smallest_eig}) we require the
following result
\begin{equation}\label{order_pdf2}
\left| {\mathbf{\Upsilon }}  \right|\prod\limits_{i < j}^s \left(
{\phi _i - \phi _j } \right) = \left| {\mathbf{\Upsilon }}  \right|
\left| \phi_i^{s-j} \right| = \sum\limits_\sigma {\sum\limits_\mu
{\operatorname{sgn} \left( \mu \right)\prod\limits_{k = 1}^s {\phi
_{\sigma _k }^{s - k} \left\{{\mathbf{\Upsilon }}  \right\}_{\sigma
_k ,\mu _k } } } },
\end{equation}
which is easily obtained using \textit{Lemma \ref{lemma1}} and the
definition of the determinant. In (\ref{order_pdf2}), $\mu  =
\left( {\mu _1 ,\mu _2 , \ldots ,\mu _s } \right)$ and $\sigma =
\left( {\sigma _1 ,\sigma _2 , \ldots ,\sigma _s } \right)$ are
permutations of $\left( {1, \ldots ,s} \right)$, the sums are over
all permutations, and ${\operatorname{sgn} \left( \cdot \right)}$
denotes the permutation sign.
Substituting (\ref{order_pdf1}) and (\ref{order_pdf2}) into
(\ref{smallest_eig}), and using \cite [Lemma 1]{Khatri1}, it can be
shown that the c.d.f.\ of $\phi _s$ can be written as
\begin{equation}\label{smallest_eig_fType}
F_{\phi _s } \left( x \right) = 1 - \Pr \left( {\phi _s  > x}
\right) = 1 - c_1 \left| {\int_x^{ + \infty } {f_i \left( {\lambda
_j ,y} \right){\rm d}y} } \right|,
\end{equation}
where
\begin{equation}\label{fi_lambdaj}
f_i \left( {\lambda _j ,y} \right) = \left\{
{\begin{array}{*{20}c}
   {y^{t - i} e^{ - y} {}_0F_1 \left( {t - s + 1;\lambda _j y} \right)} & {j = 1, \ldots ,L}  \\
   {y^{t + s - i - j} e^{ - y} \left( {t - s} \right)!/\left( {t - j} \right)!} & {j = L + 1, \ldots ,s}  \\
 \end{array} } \right.
\end{equation}
Using the relation
\begin{equation}\label{0F1_fun}
{}_0F_1 \left( {t - s + 1;x} \right) = \left( {t - s} \right)!x^{ -
\left( {t - s} \right)/2} I_{t - s} \left( {2\sqrt x } \right)
\end{equation}
as well as (\ref{nuttall_q}), the remaining integral in
(\ref{smallest_eig_fType}) can be evaluated as follows
\begin{align}\label{phiij_smallest}
\mathcal{I}_{i,j}(x) &\defeq \int_x^\infty  {f_i \left( {\lambda _j
,y}
\right){\rm d}y} \nonumber \\
&= \left\{ {\begin{array}{*{20}c}
   {\left( {t - s} \right)! \, {\lambda _j }^{\left( {s -
t} \right)/2} \, e^{\lambda _j } 2^{\left( {2i - s - t} \right)/2}
Q_{s + t - 2i + 1,t - s} \left( {\sqrt {2\lambda _j } ,\sqrt {2x} }
\right) } & {j = 1, \ldots ,L}  \\
   {\left( {t - s} \right)! \, \Gamma \left( {t + s - i - j +
1,x} \right)/\left( {t - j} \right)! } & {j = L + 1, \ldots ,s}  \\
 \end{array} } \right.
\end{align}
Since $\Pr \left( {\phi _s  \geqslant 0} \right) = 1$, $c_1$ can
also be expressed as
\begin{equation}\label{c1_express}
c_1  = 1/\left| \mathcal{I}_{i,j}(0)  \right|.
\end{equation}
Substituting (\ref{phiij_smallest}) and (\ref{c1_express}) into
(\ref{smallest_eig_fType}) and simplifying by removing common
factors from the numerator and denominator determinants, we obtain
the desired c.d.f.\ of the smallest eigenvalue.

\section{Proof of Theorem \ref{largest_cdf}}\label{sec:Proof_largest}

We can evaluate the c.d.f.\ of the maximum eigenvalue $\phi_1$ as
follows
\begin{equation}\label{largest_eig}
F_{\phi _1 } \left( x \right) = \Pr \left( {\phi _s < \ldots < \phi
_1  \leqslant x} \right)  = \int_{\mathcal{D}_2} f\left( {\phi _1 ,
\ldots ,\phi _s } \right) \, {\rm d}\phi_1 \cdots {\rm d}\phi_s \,
\end{equation}
where $D_2  = \left\{ {\phi_s  <  \ldots  < \phi_1  < x} \right\}$.
We evaluate these integrals by following a similar procedure to that
used for evaluating (\ref{smallest_eig}) in the proof of Theorem
\ref{smallest_cdf}, to obtain
%
\begin{align}\label{largest_eig_fType}
F_{\phi _1 } \left( x \right) = c_2 \left| {\int_0^x {f_i \left(
{\lambda _j ,y} \right) {\rm d}y} } \right| \; ,
\end{align}
where $f_i \left( {\lambda _j ,y} \right)$ is defined as in
(\ref{fi_lambdaj}).
We obtain the desired result by noting that $c_1 = c_2$ since $\Pr
\left( {\phi _1 < \infty } \right) = 1$, and using the property
\begin{equation}\label{integral_relation}
\int_0^x {f_i \left( {\lambda _j ,y} \right) {\rm d}y}  = \int_0^{ +
\infty } {f_i \left( {\lambda _j ,y} \right){\rm d}y}  - \int_x^{ +
\infty } {f_i \left( {\lambda _j ,y} \right) {\rm d}y}.
\end{equation}
along with (\ref{phiij_smallest}) and (\ref{c1_express}) in
(\ref{largest_eig_fType}).

\section{Proof of Theorem \ref{thekth_cdf}}\label{sec:Proof_thekthRc}

Let $D_3  = \left\{ {\phi _s  < \ldots  < x < \phi _{k - 1}  <
\ldots < \phi _1 } \right\}$, $D_4 = \left\{ {x < \phi _{k - 1}  <
\ldots  < \phi _1  <  + \infty } \right\}$ and $D_5  = \left\{ {0
< \phi _s  <  \ldots \phi _k < x} \right\}$. Using
(\ref{order_pdf1}) and (\ref{order_pdf2}) we can write the
probability $p$ as
\begin{equation}\label{p_expression1}
p = c_1 \int\limits_{D_3 } {\sum\limits_\sigma  {\sum\limits_\mu
{\operatorname{sgn} \left( \mu  \right)} } } \prod\limits_{k = 1}^s
{\phi _{\sigma _k }^{t - k} e^{ - \phi _{\sigma _k } }
\left\{{\mathbf{\Upsilon }}  \right\}_{\sigma _k ,\mu _k }} {\rm
d}\phi _k.
\end{equation}
Note that the summation over $\sigma$ can be decomposed as
\cite{Khatri1}
\begin{equation}\label{sum_equ}
\sum\limits_\sigma  { = \sum\limits_1 {\sum\nolimits_{r_{\alpha _1
} } {\sum\nolimits_{r_{\alpha _2 } } {} } } }
\end{equation}
where $\sum\nolimits_{r_{\alpha _1 } } {}$ denotes summation over
the permutations $\left( {r_{\alpha _1 } , \ldots ,r_{\alpha _{k -
1} } } \right)$ of $ \left( {1, \ldots ,k - 1} \right)$ and
$\sum\nolimits_{r_{\alpha _2 } } {}$ denotes summation over the
permutations $\left( {r_{\alpha _k } , \ldots ,r_{\alpha _s } }
\right)$ of $\left( {k, \ldots ,s} \right)$. Therefore, we have
\begin{align}\label{p_expression2}
p &=c_1 \sum\limits_1 {\sum\limits_\mu  {\int\limits_{D_3 }
{\operatorname{sgn} \left( \mu  \right)} } }
\sum\nolimits_{r_{\alpha _1 } } {\sum\nolimits_{r_{\alpha _2 } }
{\prod\limits_{k = 1}^s {\phi _{\sigma _k }^{t - k} e^{ - \phi
_{\sigma _k } } \left\{{\mathbf{\Upsilon }}  \right\}_{\sigma _k ,\mu _k }} {\rm d}\phi _k } }\nonumber\\
&= c_1\sum\limits_1 {\sum\limits_\mu  {\operatorname{sgn} \left( \mu
\right)I_1 \left( \alpha  \right)} } I_2 \left( \alpha \right),
\end{align}
where
\begin{align}\label{I1_alpha1}
I_1 \left( \alpha  \right) &= \sum\nolimits_{r_{\alpha _1 } }
{\int\limits_{D_4 } {\prod\limits_{i = 1}^{k - 1} {\phi _{r_{\alpha
_i } }^{t - \alpha _i } e^{ - \phi _{r_{\alpha _i } } } \left\{
{\mathbf{\Upsilon }}  \right\}_{r_{\alpha _i } ,j}} } } {\rm d}\phi
_{r_{\alpha _i } } = \prod\limits_{i = 1}^{k - 1} {\int_x^{ + \infty
} {\phi _{r_{\alpha _i } }^{t - \alpha _i } e^{ - \phi _{r_{\alpha
_i } } } \left\{ {\mathbf{\Upsilon }}  \right\}_{r_{\alpha _i } ,j}}
} {\rm d}\phi _{r_{\alpha _i } },
\end{align}
\begin{align}\label{I1_alpha2}
I_2 \left( \alpha  \right) &= \sum\nolimits_{r_{\alpha _2 } }
{\int\limits_{D_5 } {\prod\limits_{i = k}^s {\phi _{r_{\alpha _i }
}^{t - \alpha _i } e^{ - \phi _{r_{\alpha _i } } } \left\{
{\mathbf{\Upsilon }}  \right\}_{r_{\alpha _i } ,j}} } } {\rm d}\phi
_{r_{\alpha _i } } = \prod\limits_{i = k}^s {\int_0^x {\phi
_{r_{\alpha _i } }^{t - \alpha _i } e^{ - \phi _{r_{\alpha _i } } }
\left\{ {\mathbf{\Upsilon }}  \right\}_{r_{\alpha _i } ,j}} } {\rm
d}\phi _{r_{\alpha _i } }.
\end{align}
The last equality follows from \cite [Lemma 1]{Khatri1}. The
desired result follows from (\ref{integral_relation}),
(\ref{phiij_smallest}) and the definition of the determinant.

\section{Proof of Theorem \ref{pdf_approximation}}\label{sec:Proof_pdfappro}

Here we will derive the p.d.f.\ expansion (\ref{first_order1}).
The corresponding c.d.f.\ expansion (\ref{first_order1_cdf}) then
follows trivially.

We start by noting that since $f_{\phi _k } \left( {\phi _k }
\right) = {\rm d} F_{\phi _k } \left( {\phi _k } \right)/{\rm d}\phi
_k$, the Taylor expansion of $f_{\phi _k } \left( {\phi _k }
\right)$ around the origin can be written as
\begin{equation}\label{Taylor_expansion1}
f_{\phi _k } \left( {\phi _k } \right) = F_{\phi _k }^{\left( 1
\right)} \left( 0 \right) + F_{\phi _k }^{\left( 2 \right)} \left( 0
\right)\phi _k  +  \cdots  + \frac{{F_{\phi _k }^{\left( {q + 1}
\right)} \left( 0 \right)}} {{q!}}\phi _k^q  + o\left( {\phi _k^q }
\right).
\end{equation}
In order to simplify the derivations, we will initially work with $
F_{\phi _k }(x)$ under the assumption of full-rank $\mathbf{\Omega}$
(i.e. $L = s$). We will then generalize our result to the
arbitrary-rank $\mathbf{\Omega}$ case by evaluating limits where
necessary.

\subsection{Derivation for $\phi_1$} We first derive the first order expansion of $f_{\phi _1
} \left( {\phi _1 } \right)$. Using (\ref{largest_cdfexp}) and a
well-known result for the $k$th derivative of a determinant, we
obtain
\begin{equation}\label{det_derivative1}
{F_{\phi _1 }^{\left( {q + 1} \right)} \left( x \right)} =
\frac{1} {{\left| {{\mathbf{\Psi }}\left( 0 \right)} \right|}}
\sum\limits_{q_1  +  \cdots  + q_s  = q + 1} \frac{{(q + 1)!}} {{q_1
!q_2 ! \cdots q_s !}}
 \left| {\frac{{{\rm d}^{q_i } }}
{{{\rm d}x^{q_i } }}\left\{ {{\mathbf{\Xi }}\left( x \right)}
\right\}_{i,j} } \right| \; .
\end{equation}
%
We require the $q_i$th order derivative of the Nuttall $Q$-function
$Q_{s + t - 2i + 1,t - s} \left( {\sqrt {2\lambda _j } ,\sqrt {2x} }
\right)$ in (\ref{Xi_exp}). Omitting details, with the help of
Leibnitz' rule, we evaluate these derivatives as follows
\begin{equation}\label{Nuttall_derviate2}
Q_{s + t - 2i + 1,t - s}^{\left( {q_i} \right)} \left( {\sqrt
{2\lambda _j } ,\sqrt {2x} } \right) =  - \sum\limits_{r = 0}^{q_i -
1 } \binom{q_i-1}{r}
 g_1^{\left( {q_i - 1 - r} \right)} \left( x \right)g_2^{\left( r \right)} \left( x
 \right)
\end{equation}
for $q_i \geqslant 1$, where
\begin{equation}\label{g1x_exp}
g_1 \left( x \right) = \exp \left( { - \lambda _j  - x} \right)
\end{equation}
\begin{equation}\label{g2x_exp}
g_2 \left( x \right) = \left( {\sqrt {\lambda _j } } \right)^{t - s}
\left( {\sqrt 2 } \right)^{s + t - 2i} \sum\limits_{p = 0}^\infty
{\frac{{\lambda _j^p x^{t - i + p} }} {{p!\left( {t - s + p}
\right)!}}}.
\end{equation}
Hence,
\begin{align}\label{dnuttall_0}
&\left. {Q_{s + t - 2i + 1,t - s}^{\left( {q_i } \right)} \left( {
\sqrt {2\lambda _j } ,\sqrt {2x} } \right)} \right|_{x =
0}\nonumber\\
&{~~~~}= \left\{ {\begin{array}{*{20}c}
   0 & {q_i - 1 < t - i}  \\
   { - \left( {\sqrt {\lambda _j } } \right)^{t - s} \left( {\sqrt 2 } \right)^{s + t - 2i} \sum\limits_{r = t-i}^{q_i  - 1} {\left( {\begin{array}{*{20}c}
   {q_i  - 1}  \\
   r  \\
 \end{array} } \right)\frac{{e^{ - \lambda _j } \left( { - 1} \right)^{q_i  - 1 - r} \lambda _j^{r - t + i} r!}}
{{\left( {r - t + i} \right)!\left( {r - s + i} \right)!}}} } & {q_i - 1 \geqslant t - i}  \\
 \end{array} } \right.
\end{align}

To obtain the first order expansion of $f_{\phi _1 } \left( {\phi
_1 } \right)$, we require the minimum exponent $q$ in
(\ref{Taylor_expansion1}) (and corresponding values of $q_1,
\ldots, q_s$ in (\ref{det_derivative1})), such that
$F_{\phi_1}^{(q+1)}(0) \neq 0$. Using (\ref{det_derivative1}),
(\ref{dnuttall_0}) and the properties of the determinant, we find
that
\begin{equation}\label{ki_exp}
q_i - 1 = t - i + p_i ,
\end{equation}
where $i = 1, \ldots ,s$ and $\left( {p_1 , \ldots ,p_s } \right)$
is a permutation of $\left( {0, \ldots , s-1} \right)$. Thus
\begin{equation}\label{d1_expression}
d_1  = q = \sum\limits_{i = 1}^s {q_i }  - 1 = st - 1,
\end{equation}
and
\begin{equation}\label{delta1_expression}
a_1  = \frac{{ s t \,  e^{ - {\rm tr}\left( {\mathbf{\Lambda }}
\right)} 2^{s\left( {t - 1} \right)/2} \prod\nolimits_{i = 1}^s
{\lambda _i^{\left( {t - s} \right)/2} } \, \prod\nolimits_{i < j}^s
{\left( {\lambda _i  - \lambda _j } \right)} \, \left|
{{\mathbf{\Delta }}_1 } \right|}} {{\left| {{\mathbf{\Psi }}\left( 0
\right)} \right|\prod\nolimits_{i = 0}^{s - 1} { \left(i!\left( {t -
s + i} \right)! \right)} }},
\end{equation}
where
\begin{equation}\label{delta1_ij}
\left\{ {{\mathbf{\Delta }}_1 } \right\}_{i,j}  = 1/\left( {t + s -
i - j + 1} \right).
\end{equation}
We now simplify $a_1$. Using \cite[Eq.\ 13]{Simon} we can write
\begin{align}\label{det_psi_1}
\left| {{\mathbf{\Psi }}\left( 0 \right)} \right|
&= 2^{s\left( {t - 1} \right)/2} \left( {\prod\limits_{i = 0}^{s -
1} {i!} } \right)\left( {\prod\limits_{i = 1}^s {\lambda _i^{\left(
{t - s} \right)/2} } } \right)\left| {L_{s - i}^{\left( {t - s}
\right)} \left( { - \lambda _j } \right)} \right|.
\end{align}
We also manipulate $|{\mathbf{\Delta }}_1|$ by subtracting the
first row from all other rows, removing factors, then subtracting
the first column from all other column, and again removing
factors. This yields
\begin{align}
|{\mathbf{\Delta }}_1| = \frac{ \left((s-1)! \, (t-1)! \right)^2 }{
(t+s-1)! (t+s-2)! } | {\mathbf{\Delta} }_1^{[1,1]} |
\end{align}
where ${\mathbf{\Delta }}_1^{[1,1]}$ is the principle submatrix of
${\mathbf{\Delta }}_1$, with first row and column removed.
Continuing this same process $s-1$ more times, we obtain
\begin{align}
|{\mathbf{\Delta }}_1| = \frac{ \left( \prod_{i=1}^s (s-i)! (t-i)!
\right)^2}{ \prod_{i=1}^{2s} (t+s-i)! } = \frac{\left(\Gamma_s (s)
\Gamma_s(t) \right)^2 \Gamma_{t-s}(t-s)} {\Gamma_{t+s}(t+s)}
\label{eq:Delta1} \; .
\end{align}
Substituting (\ref{det_psi_1}) and (\ref{eq:Delta1}) into
(\ref{delta1_expression}) and simplifying yields
\begin{align}
a_1 &= \frac{ s t \, e^{-{\rm tr} (\mathbf{\Lambda}) }}{ \Gamma_s
(t+s) }
 \, \frac{ \prod\nolimits_{i < j}^s {\left( {\lambda _i  -
\lambda _j } \right)} }{ \left| L_{s-i}^{(t-s)} (-\lj) \right| } =
\frac{ s t \, e^{-{\rm tr} (\mathbf{\Lambda}) }}{ \Gamma_s (t+s) }
 \, \frac{ |
\lj^{i-1} | }{ | L_{i-1}^{(t-s)} (-\lj) | } \label{eq:a1_detfrac}
\; .
\end{align}
Finally, we remove the remaining determinant ratio. To do this, we
manipulate the numerator determinant in such a way as to construct
a scaled version of the denominator determinant. Start by
considering the row $i = s$.  Using (\ref{laguerre_k}), the
elements in this row of the denominator determinant can be written
as
\begin{align} \label{eq:Lex}
L_{s-1}^{(t-s)} (-\lj) = \sum_{k=0}^{s-2} \binom{t-1}{i-1-k} \frac{
\lj^k }{k!} + \frac{ \lj^{(s-1)} }{(s-1)!}, \; \hspace*{1cm} j = 1,
\ldots, s
\end{align}
We construct this row as the last row of the numerator determinant
by first dividing the $s$th row in the numerator determinant by
$(s-1)!$, and multiplying the determinant by $(s-1)!$ to compensate.
This gives elements corresponding to the $(s-1)^{{\rm th}}$ order
polynomial terms on the right-hand side of (\ref{eq:Lex}). All of
the remaining terms on the right-hand side of (\ref{eq:Lex}) can be
generated from weighted sums of the rows $i = 1, \ldots, s-1$ in the
numerator determinant (these row operations do not change the value
of the determinant). If we then apply the same process to rows $i =
s-1, \ldots, 1$, in order, at each stage constructing another row of
the denominator determinant, and pulling out a multiplicative factor
of $(i-1)!$, we obtain
\begin{align} \label{eq:detfrac_simple}
\frac{ | \lj^{i-1} | }{ | L_{i-1}^{(t-s)} (-\lj) | } = \prod_{i=1}^s
(s-i)! = \Gamma_s (s) \; .
\end{align}
Substituting (\ref{eq:detfrac_simple}) into (\ref{eq:a1_detfrac})
gives the desired result for $a_1$.  Note that in this case, $a_1$
is only a function of ${\rm tr}(\mathbf{\Omega})$, and is
independent of the rank of $\mathbf{\Omega}$.

\subsection{Derivation for $\phi_k \, (k = 2, \ldots, s)$}

Now consider the case $k > 1$.
Starting with (\ref{thekthcdf_exp}) for the case of full-rank
$\mathbf{\Omega}$, We observe that the minimum exponent $q$ is
obtained when $\left( {\alpha _1 , \ldots ,\alpha _s } \right) =
\left( {1, \ldots ,s} \right)$. Applying the same steps as for the
$\phi_1$ case, we obtain
\begin{equation}\label{pdf_approk1}
q_i  = \left\{ {\begin{array}{*{20}c}
   0 & {i = 1, \ldots ,k - 1}  \\
   {t - i + e_i + 1} & {i = k, \ldots ,s}  \\
 \end{array} } \right.
\end{equation}
where $\left( {e_k , \ldots ,e_s } \right)$ is a permutation of
$\left( {0, \ldots s - k} \right)$. Hence
\begin{equation}\label{dk_expression}
d_k  = q - 1 = \sum\limits_{i = k}^s {q_i } - 1 = s_k t_k - 1,
\hspace*{1.5cm} 2 \leqslant k \leqslant s.
\end{equation}
Combining (\ref{d1_expression}) and (\ref{dk_expression}) yields
(\ref{dk_totalexp}). Also, we obtain
\begin{align} \label{eq:ak}
a_k = \frac{ s_k t_k c_3 \, 2^{s_k (t_k-1)/2} |\mathbf{\Delta}_2|
|\mathbf{\Delta}_3|}{ \Gamma_{s_k}(s_k) \Gamma_{s_k}(t_k)  }
\end{align}
where
\begin{equation}\label{delta2_ijexp}
\left\{ {{\mathbf{\Delta }}_2 } \right\}_{i,j}  = \left\{
{\begin{array}{*{20}c}
   {Q_{s + t - 2i + 1,t - s} \left( {\sqrt {2\lambda _j } ,0} \right)} & {i = 1, \ldots ,k - 1}  \\
   \left( {\sqrt {\lambda _j } } \right)^{t + s - 2i} e^{ - \lambda _j
} & {i = k, \ldots ,s}  \\
 \end{array} } \right.
\end{equation}
and
\begin{equation}\label{delta3_ijexp}
\left\{ {{\mathbf{\Delta }}_3 } \right\}_{i,j}  = 1/\left( {t_k +
s_k - i - j + 1} \right) ~~~{\text{  }}i,j = 1, \ldots ,s - k + 1.
\end{equation}
Now, using \cite[Eq.\ 13]{Simon}, we factorize $|\mathbf{\Delta}_2|$
as follows
\begin{align} \label{eq:Delta2}
| \mathbf{\Delta}_2 | = \left( \prod_{j=1}^s \lj^{(t-s)/2} \right)
\Gamma_{k-1}(s) \, 2^{(-k^2+k(t+s+1)-(t+s))/2} \, |\mathbf{\Delta}_4
|
\end{align}
where
\begin{align}
\left\{ {{\mathbf{\Delta }}_4 } \right\}_{i,j}  = \left\{
{\begin{array}{*{20}c}
   \sum_{\ell=0}^{s-i} \binom{t-i}{s-i-\ell} \lambda_j^\ell / \ell!  & {i = 1, \ldots ,k - 1}  \\
   { \lambda_j^{s-i} e^{ - \lambda _j
}} & {i = k, \ldots ,s}  \\
 \end{array} } \right.
\end{align}
We evaluate $|\mathbf{\Delta}_3|$ using (\ref{delta1_ij}) and
(\ref{eq:Delta1}) as
\begin{align} \label{eq:Delta3}
|\mathbf{\Delta}_3| &= \frac{\left(\Gamma_{s_k} (s_k)
\Gamma_{s_k}(t_k) \right)^2 \Gamma_{t_k-s_k}(t_k-s_k)}
{\Gamma_{t_k+s_k}(t_k+s_k)} \; .
\end{align}
Now substituting (\ref{eq:Delta3}) and (\ref{eq:Delta2}) into
(\ref{eq:ak}) and simplifying yields
\begin{align} \label{eq:ak_simple}
a_k = \frac{ s_k t_k \Gamma_{k-1}(s) \Gamma_{s_k}(s_k)
}{\Gamma_{s_k}(t_k+s_k)} \frac{ | {\mathbf{\Delta }_4} |}{
\prod\nolimits_{i < j}^s {\left( {\lambda _i  - \lambda _j }
\right)}}
\end{align}
Finally, we generalize this result to arbitrary-rank
$\mathbf{\Omega}$.  To this end, we require the following limit
\begin{align}
\lim_{\lambda_{L+1} \rightarrow 0, \ldots, \lambda_s \rightarrow 0}
\frac{ | \mathbf{\Delta}_4 | }{ \prod\nolimits_{i < j}^s {\left(
{\lambda _i  - \lambda _j } \right)} } ,
\end{align}
which we easily evaluate using (\ref{df1_det_lim1}), and substitute
into (\ref{eq:ak_simple}) to obtain the final expression.

\newpage
\begin{figure}
\centering
\includegraphics[scale=0.7]{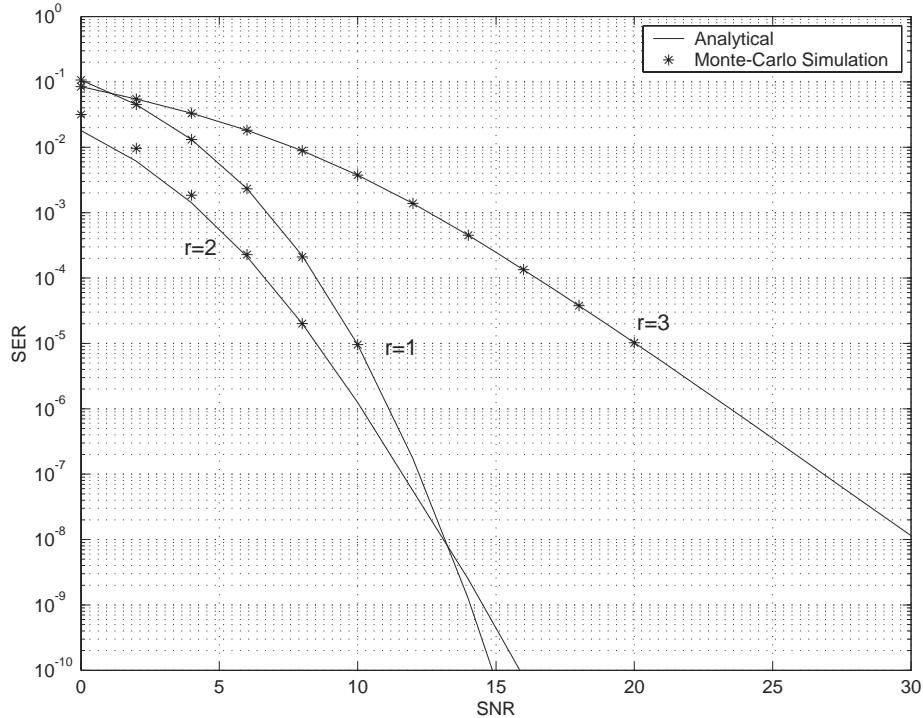}
\caption{SER for a $3 \times 5$ MB MIMO system with different
numbers of active subchannels, in Ricean fading with rank-3 mean
matrix and $K=0 {\rm dB}$.  Spectral efficiency is $3$ bits/s/Hz. }
\label{fig:fig5}
\end{figure}
\begin{figure}
\centering
\includegraphics[scale=0.7]{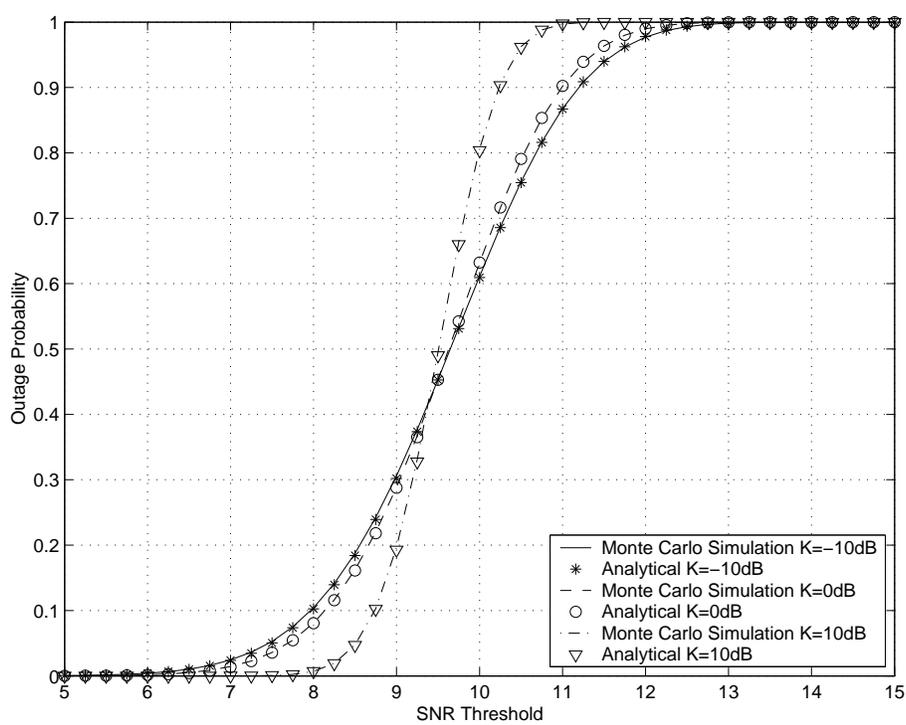}
\caption{Outage probability of $3 \times 5$ MIMO-MRC in Ricean
channels with rank-3 mean matrix, various $K$-factors, and for $P =
0{\rm dB}$.} \label{fig:fig4}
\end{figure}

\end{document}